\documentclass[twocolumn,aps,prc,superscriptaddress,showpacs,floatfix,longbibliography]{revtex4-1}
\usepackage{url}
\usepackage{cancel}
\usepackage[colorlinks,linkcolor=blue,citecolor=blue,filecolor=black,urlcolor=blue]{hyperref}
\usepackage{epsfig,graphics}
\usepackage{graphicx}
\usepackage{dcolumn}
\usepackage{bm}
\usepackage[usenames]{color}
\usepackage{amssymb}
\usepackage{amsmath}
\usepackage{multirow}
\usepackage{float}
\usepackage{harpoon}
\usepackage{MnSymbol}
\usepackage{appendix}
\usepackage{color}
\usepackage{hyperref}
\usepackage{cleveref}
\usepackage{lineno}

\begin{document}

\title{Revisiting angular momentum conservation in transport simulations of intermediate-energy heavy-ion collisions}
\author{Rong-Jun Liu}
\affiliation{Shanghai Institute of Applied Physics, Chinese Academy of Sciences, Shanghai 201800, China}
\affiliation{University of Chinese Academy of Sciences, Beijing 100049, China}
\author{Jun Xu}\email[Correspond to\ ]{xujun@zjlab.org.cn}
\affiliation{School of Physics Science and Engineering, Tongji University, Shanghai 200092, China}
\affiliation{Shanghai Advanced Research Institute, Chinese Academy of Sciences, Shanghai 201210, China}
\affiliation{Shanghai Institute of Applied Physics, Chinese Academy of Sciences, Shanghai 201800, China}

\begin{abstract}
Based on the well-calibrated IBUU transport model, we have studied the dynamical effect of incorporating rigorous angular momentum conservation in each collision of particles with homework setups. The constraint of the rigorous angular momentum conservation requires in-plane collisions and side jumps of particles after their collision. Since the option is not unique, we have compared two typical prescriptions with the original one. While the results depend quantitatively on the choice of the prescription, we found that the angular momentum conservation generally reduces local density fluctuations and thus the collision rate, and may have some influence on the density evolution, the collective flow, and even the pion production in transport simulations of intermediate-energy heavy-ion collisions.
\end{abstract}
\maketitle

\section{Introduction}
\label{sec:intro}

Transport models are among the best tools for describing the non-equilibrium dynamics in heavy-ion collisions~\cite{Xu:2019hqg,TMEP:2022xjg}. While one of the main purposes of intermediate-energy heavy-ion collisions is to extract the soft mean-field potential by comparing experimental data with results from transport simulations, hard collisions become increasingly important and dominate the dynamics in heavy-ion collisions at higher energies. For different prescriptions on nucleon-nucleon collisions in transport models, we refer the reader to Ref.~\cite{Zhang:2017esm}. However, the angular momentum conservation in each nucleon-nucleon collision is mostly neglected in these prescriptions, since its effect on dynamics was found to be small as first reported in Ref.~\cite{Gale:1990zz}.

The importance of the angular momentum conservation was recently recalled in order to explain the ``sign problem'' of the local spin polarization in relativistic heavy-ion collisions~\cite{PhysRevLett.125.062301}. It has been found that the global spin polarization of $\Lambda$ hyperons perpendicular to the reaction plane, which can be measured experimentally through the angular distribution of their weak decays~\cite{STAR:2017ckg}, can be well described with the assumption that the spin polarization is thermalized by the vortical field~\cite{PhysRevC.95.054902} produced in non-central heavy-ion collisions. On the other hand, the thermal model predicts an opposite azimuthal angular dependence of the local spin polarization in the beam direction compared to the experimental data~\cite{PhysRevLett.120.012302,PhysRevLett.123.132301}, leading to the so-called ``sign problem". While a recently improved thermal model with thermal shear could be able to explain the measured local spin polarization~\cite{Becattini:2021suc,Fu:2021pok}, it is remarkable to see that experimental data can be reproduced by a chiral transport model with rigorous angular momentum conservation~\cite{PhysRevLett.125.062301}.

In the present study, we revisit in detail how the constraint of the angular momentum conservation in the collision prescription affects the dynamics of intermediate-energy heavy-ion collisions from transport simulations in the absence of spin degree of freedom. Naively, one expects that the final momenta of particles after their collision should be in the same plane as their initial momenta in the center-of-mass (C.M.) frame in order to conserve the direction of the angular momentum, so this changes the azimuthal angular dependence of each collision. To conserve the magnitude of the angular momentum, side jumps of particles after their collision are generally needed~\cite{PhysRevLett.113.182302,PhysRevLett.115.021601}. Both may have some influence on the dynamics of the heavy-ion simulation. In the present baseline calculation, we use homework setups for the mean-field potential and collision cross sections, and try to incorporate the constraint of the angular momentum conservation into elastic and inelastic collisions. We find that there are different options for the collision prescriptions that conserve angular momentum, and they may affect the final nucleon observables as well as the pion production.

\section{Theoretical framework}
\label{sec:theory}

The simulation is carried out based on the isospin-dependent Boltzmann-Uheling-Uhlenbeck (IBUU) transport model, and details of the code can be found in Ref.~\cite{TMEP:2022xjg}. This model has been well calibrated by the previous efforts of the transport model evaluation project~\cite{Xu:2016lue,Zhang:2017esm,Ono:2019ndq,Colonna:2021xuh}. In this section, we briefly describe the basic setups of the transport simulation in the present study, and will mainly focus on the collision treatment. To incorporate the constraint of the angular momentum conservation, we develop two collision prescriptions, and will compare them with the original one.

\subsection{Basic setups of transport simulation}

We use simplified but reasonable setups in the IBUU transport model to illustrate the effect of the angular momentum conservation on the dynamics of intermediate-energy heavy-ion collisions. The simulations are mainly focused on non-central Au+Au collisions at different beam energies. The initial coordinates of neutrons and protons are sampled according to their density distributions in Au nucleus obtained based on the Skyrme-Hartree-Fock model with the MSL0 force~\cite{Chen:2010qx}, and their initial momenta are sampled isotropically within the local isospin-dependent Fermi sphere. The nucleon momenta are then boosted according to the beam energy, and the simulation is performed in the C.M. frame of Au+Au collisions. We use the Skyrme-like momentum-independent mean-field potential including a symmetry potential linear to the density, which reproduces the empirical nuclear matter properties, and their detailed forms are the same as those used in Ref.~\cite{Xu:2016lue}. To implement the mean-field potential, we employ the lattice Hamiltonian method~\cite{Lenk:1989zz}. We use point particles in the implementation of the Coulomb force, for which the cut-off distance is set to be 1 fm to avoid divergence.

We use the modified Bertsch's prescription~\cite{Bertsch:1988ik,Zhang:2017esm} for nucleon-nucleon collisions. The minimum distance of two colliding particles in their C.M. frame perpendicular to their relative velocity is
\begin{equation}
{d_\perp^\star}^2 = (\vec{r}_1^\star-\vec{r}_2^\star)^2 - \frac{[(\vec{r}_1^\star-\vec{r}_2^\star)\cdot \vec{v}_{12}^\star]^2}{{v_{12}^\star}^2},
\end{equation}
where $\vec{r}_1^\star$ and $\vec{r}_2^\star$ are positions of the two particles, and $\vec{v}_{12}^\star = \vec{v}_1^\star - \vec{v}_2^\star$ is their relative velocity, with the asterisk representing the quantity in the C.M. frame of the colliding particles. The collision can happen if the condition
\begin{equation}
\pi {d_\perp^\star}^2 < \sigma
\end{equation}
is satisfied, and we use a constant and isotropic baryon-baryon elastic cross section $\sigma=40$ mb in the present study. Whether the collision happens in this time step is determined by the condition of the closest approach, i.e.,
\begin{equation}
|(\vec{r}_1^\star-\vec{r}_2^\star)\cdot \vec{v}_{12}^\star/{v_{12}^\star}^2|< \frac{1}{2} \delta t.
\end{equation}
We set $\delta t = \Delta t/\gamma$, where $\gamma=1/\sqrt{1-\vec{\beta}^2}$ is the Lorentz factor with $\vec{\beta}$ being the average velocity of the colliding pair in the computational frame. We have also removed the spurious collisions by setting that the two particles, which have collided once, can not collide again unless one of them has collided with a third particle. We use point nucleons in the implementation of the Pauli blocking, where the cubic local phase-space occupation probability with the dimension $\Delta r=2$ fm and $\Delta p=0.1$ GeV/c is calculated to evaluate the Pauli blocking probability together with the interpretation method. If the collision is Pauli blocked, both the momenta and coordinates of the colliding particles are retained. Besides elastic $N+N \rightarrow N+N$ collisions, we have also incorporated the channels for the pion production, including $N+N \leftrightarrow N+\Delta$ and $\Delta \leftrightarrow N+\pi$ channels for different isospin states of nucleons, $\Delta$ resonances, and pions. The detailed inelastic cross sections, decay width, and $\Delta$ and pion masses are exactly the same as those in the homework setups of Ref.~\cite{Ono:2019ndq}.

\subsection{Collision prescriptions}

For the ease of discussion on the collision prescription for particle 1 and particle 2, we first give the basic quantities and their relations in their C.M. (collision) frame and computational (heavy-ion simulation) frame. The momentum $\vec{p}_{1}^\star$, the coordinate $\vec{r}_{1}^\star$, the time $t_{1}^\star$, and the energy $E_{1}^\star$ of particle $1$ before collision in the C.M. frame can be expressed in terms of those in the computational frame as follows
\begin{eqnarray}
\vec{p}_{1}^\star&=&\vec{p}_1+a_{1}\vec{\beta}, \notag\\
\vec{r}_{1}^\star&=&\vec{r}_1+b_{1}\vec{\beta}, \notag\\
t_{1}^\star&=&\gamma(t_{1}-\vec{r}_{1}\cdot\vec{\beta}), \notag\\
E_{1}^\star&=&\gamma(E_{1}-\vec{p}_{1}\cdot\vec{\beta}), \notag
\end{eqnarray}
with $a_{1}=\gamma(\frac{\gamma}{\gamma+1}\vec{p}_{1}\cdot\vec{\beta}-E_{1})$ and $b_{1}=\gamma(\frac{\gamma}{\gamma+1}\vec{r}_{1}\cdot\vec{\beta}-t_{1})$. For particle 2, the same relations hold with the subscript $1 \rightarrow 2$. In the C.M. frame, the momenta satisfy $\vec{p}_{1}^\star=-\vec{p}_{2}^\star \equiv \vec{p}^\star$. We assume that the collision occurs at the same time ($t_1=t_2=0$) in the computational frame, but generally at different times in the C.M. frame ($t_{1}^\star \neq t_{2}^\star$). Similarly, the corresponding quantities in the computational frame after collision can be expressed in terms of those in the C.M. frame as
\begin{eqnarray}
\vec{p'}_{1}&=&\vec{p'}_{1}^\star+{a'}_{1}^\star\vec{\beta}, \notag\\
\vec{r'}_{1}&=&\vec{r'}_{1}^\star+{b'}_{1}^\star\vec{\beta}, \notag\\
t'_{1}&=&\gamma({t'}_{1}^\star+\vec{r'}_{1}^\star\cdot\vec{\beta}), \notag\\
E'_{1}&=&\gamma({E'}_{1}^\star+\vec{p'}_{1}^\star\cdot\vec{\beta}), \notag
\end{eqnarray}
with ${a'}_{1}^\star=\gamma(\frac{\gamma}{\gamma+1}\vec{p'}_{1}^\star\cdot\vec{\beta}+{E'}_{1}^\star)$ and ${b'}_1^\star=\gamma(\frac{\gamma}{\gamma+1}\vec{r'}_{1}^\star\cdot\vec{\beta}+{t'}_{1}^\star)$. As is seen, quantities with a prime represent those after collisions. Similarly, $\vec{p'}_{1}^\star=-\vec{p'}_{2}^\star \equiv \vec{p'}^\star$ is satisfied. In order to set the same time $t'_1=t'_2=0$ for particles after collision in the computational frame, we set ${t'}_1^\star=-\vec{r'}^\star_{1}\cdot\vec{\beta}$ and ${t'}_2^\star=-\vec{r'}_{2}^\star\cdot\vec{\beta}$.

\begin{figure}[h]
\includegraphics[width=0.5\linewidth]{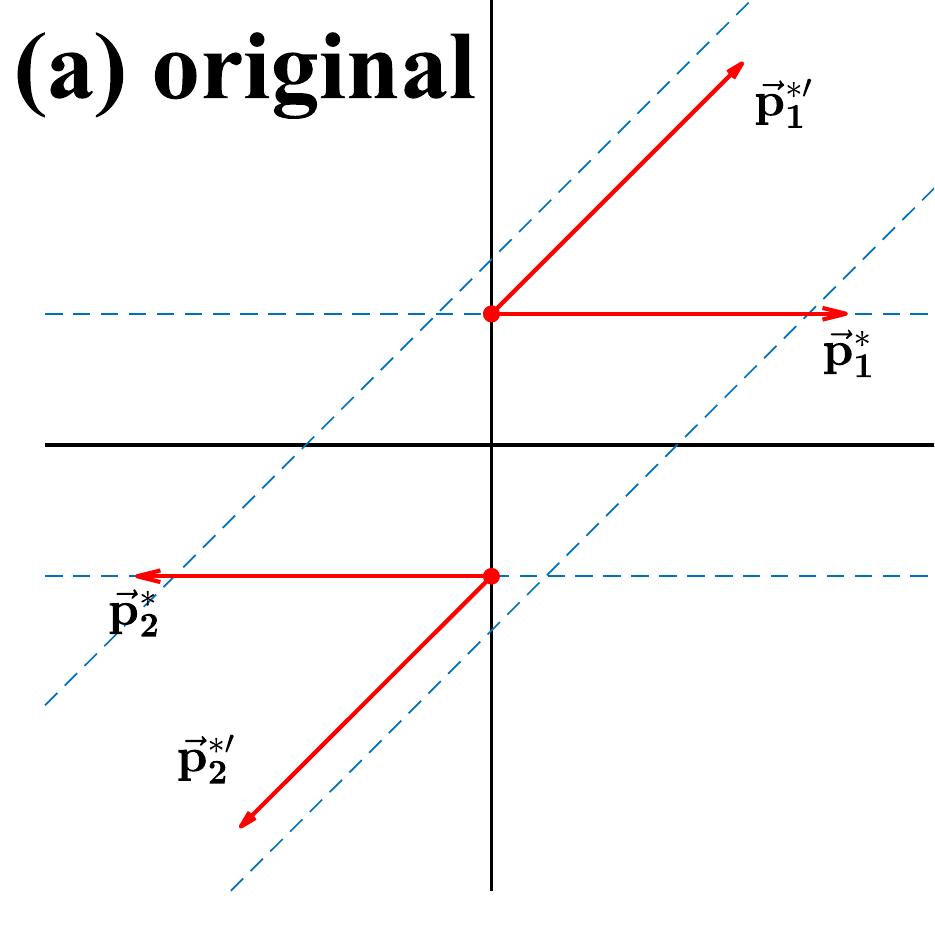}\\
\includegraphics[width=0.5\linewidth]{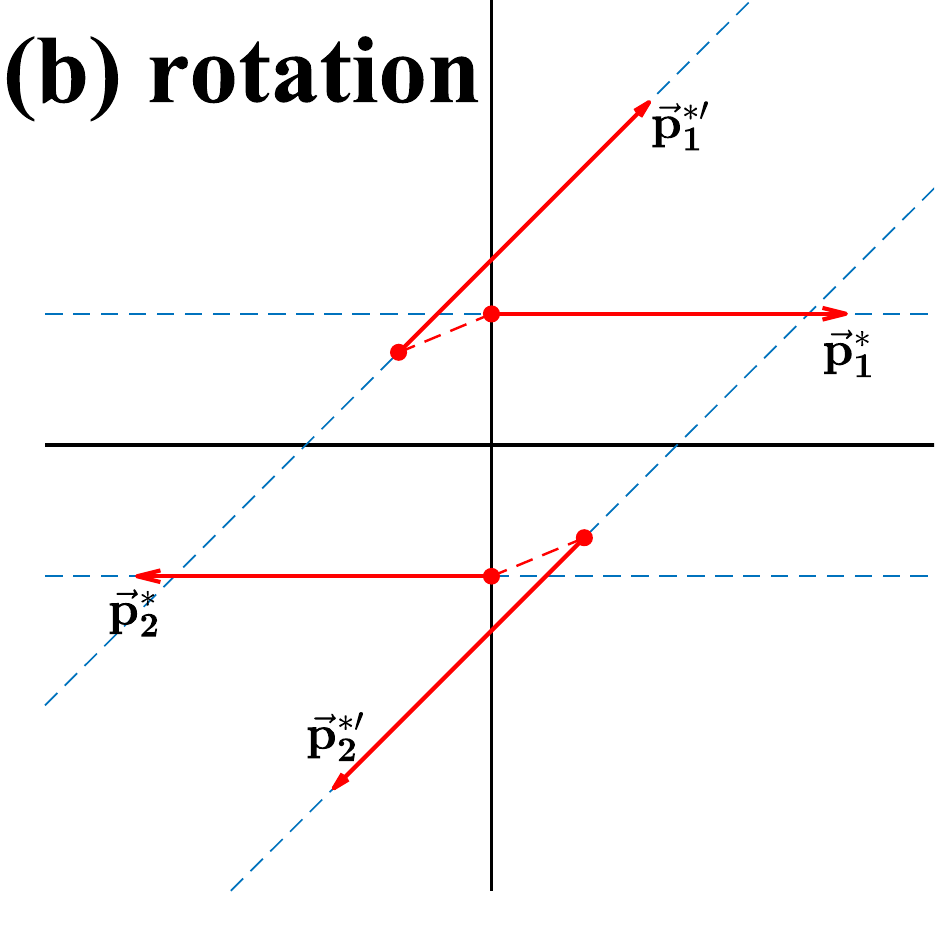}\\
\includegraphics[width=0.5\linewidth]{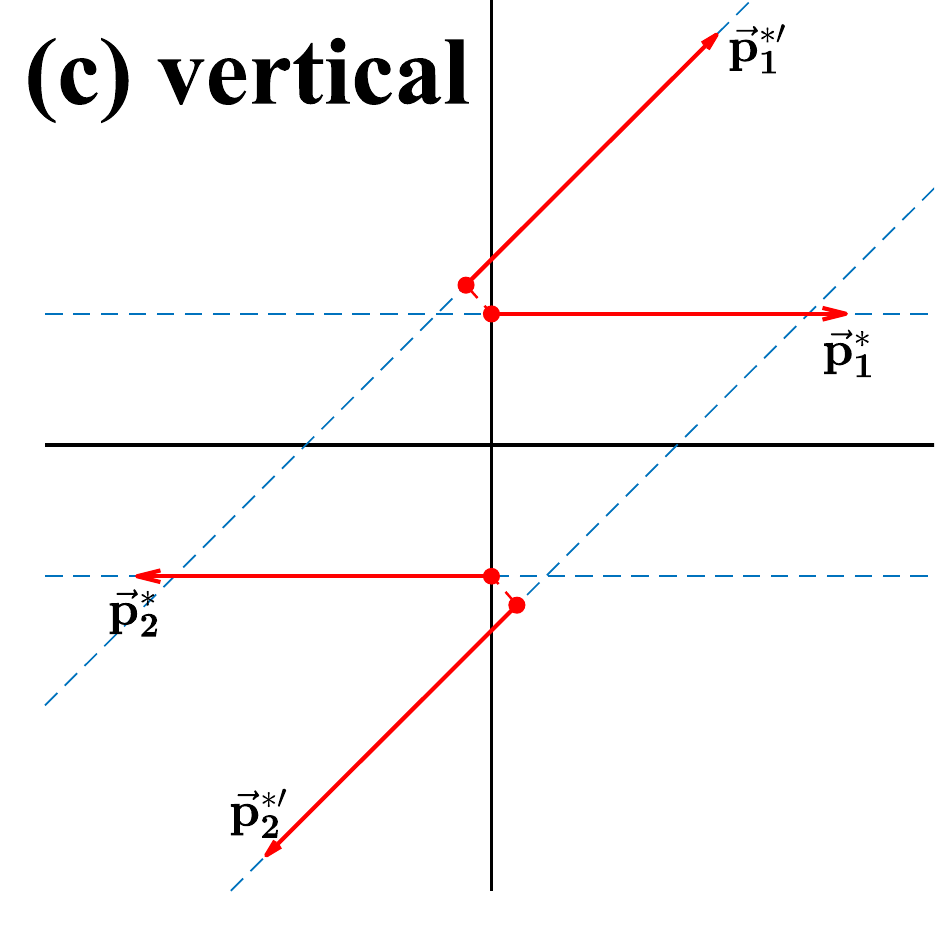}
\caption{\label{cartoon} Cartoons for three different prescriptions in the C.M. frame of the collision. }
\end{figure}

Defining the average and relative coordinates before collision in the C.M. frame as
\begin{eqnarray}
\vec{R}^\star = (\vec{r}_1^\star+\vec{r}_2^\star)/2, ~~~ \vec{r}^\star = (\vec{r}_1^\star-\vec{r}_2^\star)/2, \notag
\end{eqnarray}
the angular momentum before collision in the computational frame can be expressed as
\begin{eqnarray}
    \vec{J}&=&\vec{r}_{1}\times\vec{p}_{1}+\vec{r}_{2}\times\vec{p}_{2} \notag \\
    &=&(\vec{r}_{1}^\star+b_1^\star\vec{\beta})\times(\vec{p}_{1}^\star+a_1^\star\vec{\beta})+(\vec{r}_{2}^\star+b_2^\star\vec{\beta})\times(\vec{p}_{2}^\star+a_2^\star\vec{\beta}) \notag \\
    &=&\vec{J}^\star+(b_1^\star-b_2^\star)\vec{\beta}\times\vec{p}^\star+(a_1^\star+a_2^\star)\vec{R}^\star\times\vec{\beta}+(a_1^\star-a_2^\star)\vec{r}^\star\times\vec{\beta} \notag \\
    &=&\vec{J}^\star+\gamma(t_{1}^\star-t_{2}^\star)\vec{\beta}\times\vec{p}^\star+\gamma{E^\star}\vec{R}^\star\times\vec{\beta}+\gamma(E_{1}^\star-E_{2}^\star)\vec{r}^\star\times\vec{\beta} \notag \\
    &=&\vec{J}^\star+[-\gamma(t_{1}^\star-t_{2}^\star)\vec{p}^\star+\gamma{E^\star}\vec{R}^\star+\gamma(E_{1}^\star-E_{2}^\star)\vec{r}^\star]\times\vec{\beta}, \label{J1}
\end{eqnarray}
and the angular momentum after collision in the computational frame can be expressed similarly as
\begin{equation}
\vec{J'}=\vec{J'}^\star+[-\gamma({t'}_{1}^\star-{t'}_{2}^\star)\vec{p'}^\star+\gamma{{E'}^\star}\vec{R'}+\gamma({E'}_{1}^\star-{E'}_{2}^\star)\vec{r'}^\star]\times\vec{\beta}. \label{J2}
\end{equation}
In the above, $E^\star={E'}^\star=E_1^\star+E_2^\star={E'}_1^\star+{E'}_2^\star$ is satisfied from the energy conservation condition. $\vec{J}^\star=2\vec{r}^\star\times\vec{p}^\star$ and $\vec{J'}^\star=2\vec{r'}^\star\times\vec{p'}^\star$ are the angular momentum in the C.M. frame before and after collision, respectively. To assure the conservation of the angular momentum in the computational frame ($\vec{J}=\vec{J'}$), we require the angular momentum conservation in the C.M. frame ($\vec{J}^\star=\vec{J'}^\star$) as well as that from the C.M. motion, i.e., the rest parts in Eqs.~(\ref{J1}) and (\ref{J2}) should be equal. The latter requires a shift of the C.M. coordinate
\begin{eqnarray}
    \Delta \vec{R}^\star&=&\frac{1}{E^\star}[({t'}_{1}^\star-{t'}_{2}^\star)\vec{p'}^\star-(t_{1}^\star-t_{2}^\star)\vec{p}^\star \notag\\
    &-&({E'}_{1}^\star-{E'}_{2}^\star)\vec{r'}^\star+(E_{1}^\star-E_{2}^\star)\vec{r}^\star].
\end{eqnarray}
Therefore, the coordinates of particle 1 and particle 2 after collision in their C.M. frame can be respectively written as
\begin{eqnarray}
\vec{r'}_{1}^\star&=&\Delta \vec{R}^\star+\vec{R}^\star+\vec{r'}^\star, \\
\vec{r'}_{2}^\star&=&\Delta \vec{R}^\star+\vec{R}^\star-\vec{r'}^\star,
\end{eqnarray}
representing the so-called ``side jump''.

We illustrate with Fig.~\ref{cartoon} how to conserve the angular momentum ($\vec{J}^\star=\vec{J'}^\star$) in the C.M. frame of an elastic collision. In the original collision prescription, as shown in Fig.~\ref{cartoon} (a), the coordinates of colliding particles are unchanged after collision, and their momenta are sampled in the $4\pi$ solid angle according to the differential cross section. In this way, both the direction and the magnitude of the angular momentum in the C.M. frame of the collision are generally changed. To conserve the angular momentum, the coordinates of the particles after collision should generally be changed, once the directions of their momenta after collision are determined. Given $|\vec{p}_1^\star|=|\vec{p'}_1^\star|=|\vec{p}_2^\star|=|\vec{p'}_2^\star|$, the constraint of the angular momentum conservation requires an in-plane collision, i.e., without changing the azimuthal angle, and the distance between the horizontal blue dashed lines, representing the direction of the initial momenta, should be the same as that between the inclined blue dashed lines, representing the directions of the final momenta. Then, there are infinite options to achieve that, and in the present study we consider two typical options. Figure~\ref{cartoon} (b) shows a similar collision prescription as in Ref.~\cite{PhysRevLett.125.062301}, where both the coordinates and the momenta of the colliding particles rotate around their C.M. with a certain angle, and we dub this prescription as ``rotation'' in the following discussion. Figure~\ref{cartoon} (c) shows a collision prescription that the coordinates of particles after collision move vertically to the blue dashed line, and we dub it as ``vertical'', which is identical to the prescription in Ref.~\cite{Gale:1990zz}. In the ``vertical'' prescription, we have the least change of the coordinates after collision in the C.M. frame, so in principle this prescription should be most similar to the original one, except that an in-plane collision is required.

The above prescription for elastic collisions can be easily generalized to inelastic collisions. For the $N+N \leftrightarrow N+\Delta$ channel in the present study, we have $|\vec{p}^\star|\neq |\vec{p'}^\star|$, and the distance between the inclined blue dashed lines should be modified to satisfy $\vec{r}^\star\times\vec{p}^\star=\vec{r'}^\star\times\vec{p'}^\star$. In the $\Delta \rightarrow N+\pi$ process, $N$ and $\pi$ are produced at the same position as $\Delta$, and the angular momentum is always conserved. We are unable to conserve the angular momentum in the $N+\pi \rightarrow \Delta$ process without further incorporating spin degree of freedom, and leave this channel as it is in the original IBUU model.

\section{Results and discussions}
\label{sec:results}

We now compare in details simulation results from the three collision prescriptions as shown in Fig.~\ref{cartoon}. We first present basic tests on the collision rate in a box calculation, compare the total angular momentum in non-central Au+Au collisions, and display the distribution of the side-jump distance for colliding particles after collisions. We then compare in detail the nucleon dynamics and nucleon observables as well as the pion production from the three collision prescriptions in non-central Au+Au collisions.

\subsection{Basic tests}

\begin{figure}[ht]
\includegraphics[width=1\linewidth]{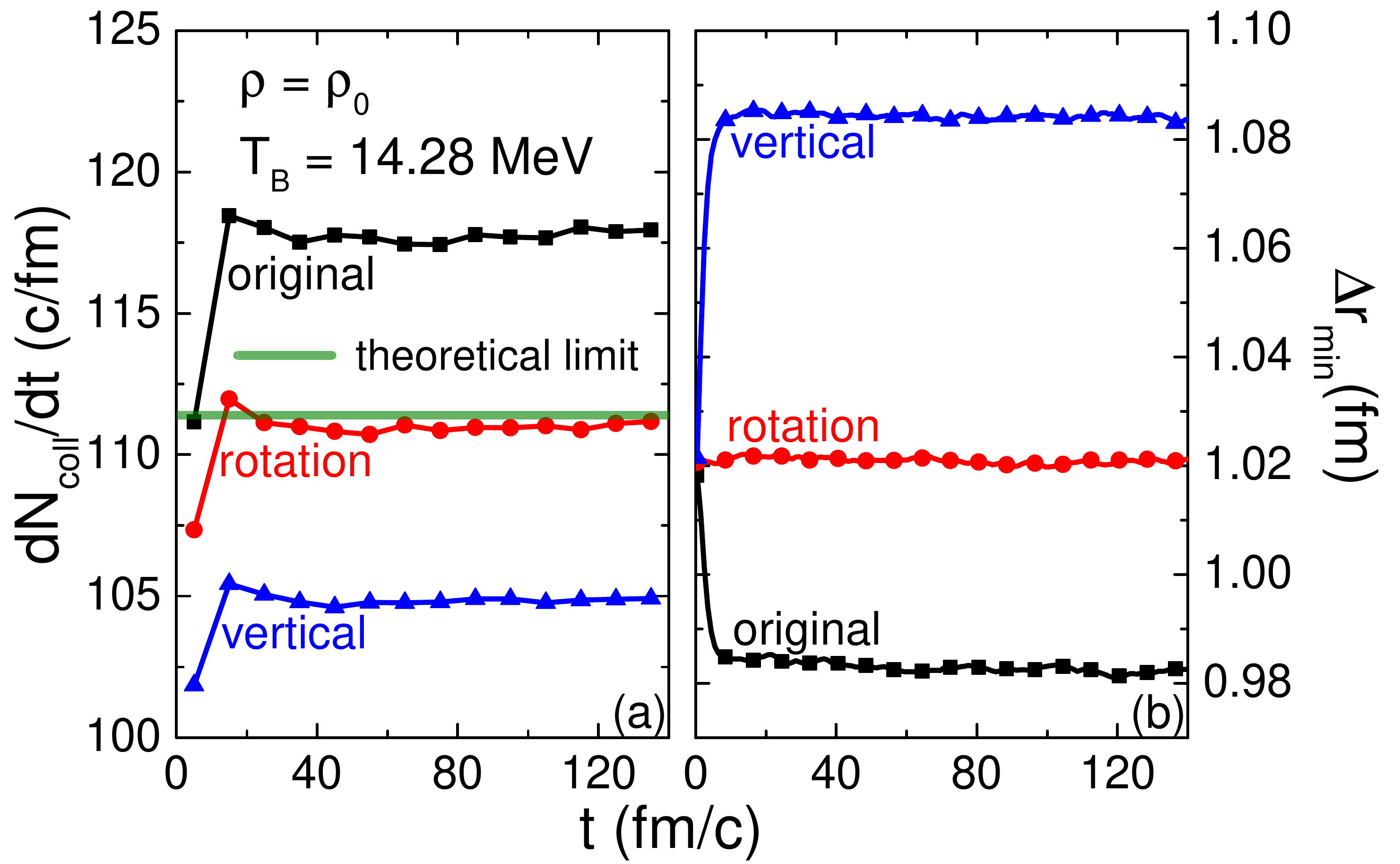}
\caption{\label{dNdt} Left: Nucleon-nucleon elastic collision rates from three collision prescriptions in the box calculation compared with the theoretical limit; Right: Event average values of the neighboring distance for each nucleon from three collision prescriptions in the box calculation. }
\end{figure}

Before simulating heavy-ion collisions, we first check with the elastic nucleon-nucleon collision rate in a box with the periodic boundary condition. The box has a size of $20\times20\times20$ fm$^3$. Initial coordinates of 640 neutrons and 640 protons are sampled uniformly within the box, and their initial momenta are sampled within the Fermi sphere. Thus, the system is initialized at zero temperature and at the saturation density. Without incorporating the mean-field potential, Pauli blocking, or Coulomb potential, the system evolves with only elastic nucleon-nucleon collisions, and the momentum distribution gradually changes to a Boltzmann distribution at the temperature of about $T=14.28$ MeV as a result of energy conservation, as in Ref.~\cite{Zhang:2017esm}. Figures~\ref{dNdt} (a) compares the collision rates from the three collision prescriptions as in Fig.~\ref{cartoon} with the theoretical limit $dN_{coll}/dt=111.4$ c/fm. The increase of the collision rate at early stage represents the slightly higher collision rate from the Boltzmann distribution than the initial Fermi-Dirac distribution. The ``original'' collision prescription overestimates the theoretical limit, as shown in Ref.~\cite{Zhang:2017esm}, as a result of higher-order correlations. The ``rotation'' prescription leads to a collision rate consistent with the theoretical limit, while the ``vertical'' prescription underestimates the theoretical limit. On the other hand, it is remarkable to see that the new prescriptions reproduce the theoretical limit within $5\%$. We have checked that the three prescriptions lead to exactly the same momentum distribution after reaching thermal equilibrium. To understand the different collision rate, we have further checked the distribution of nucleon coordinates. While the coordinates are generally uniformly distributed within the box, there are always local density fluctuations, and different collision prescriptions may lead to different local density fluctuations and thus different collision rates. We use the event average value of the neighboring distance for each nucleon $\Delta r_{min} = \sum_{i=1}^N \min\{|\vec{r_i}-\vec{r_j}|\}/N$, where particle $j$ is in the same event of particle $i$, and $N=1280$ is the total nucleon number in each event, to qualify the local density fluctuation. As shown in Fig.~\ref{dNdt} (b), the initial value of $\Delta r_{min}$ is about 1.02 fm, while it soon becomes stable at different values for different collision prescriptions. Prescriptions of in-plane collisions, especially the ``vertical'' prescription which has the minimum change of the coordinate after collision, lead to larger $\Delta r_{min}$, thus weaker local density fluctuations and lower collision rates. We note that for a simple and face-centered cubic lattice, $\Delta r_{min}$ can be as large as 1.842 and 2.067 fm at the saturation density in a box system, respectively.

\begin{figure}[h]
\includegraphics[width=1\linewidth]{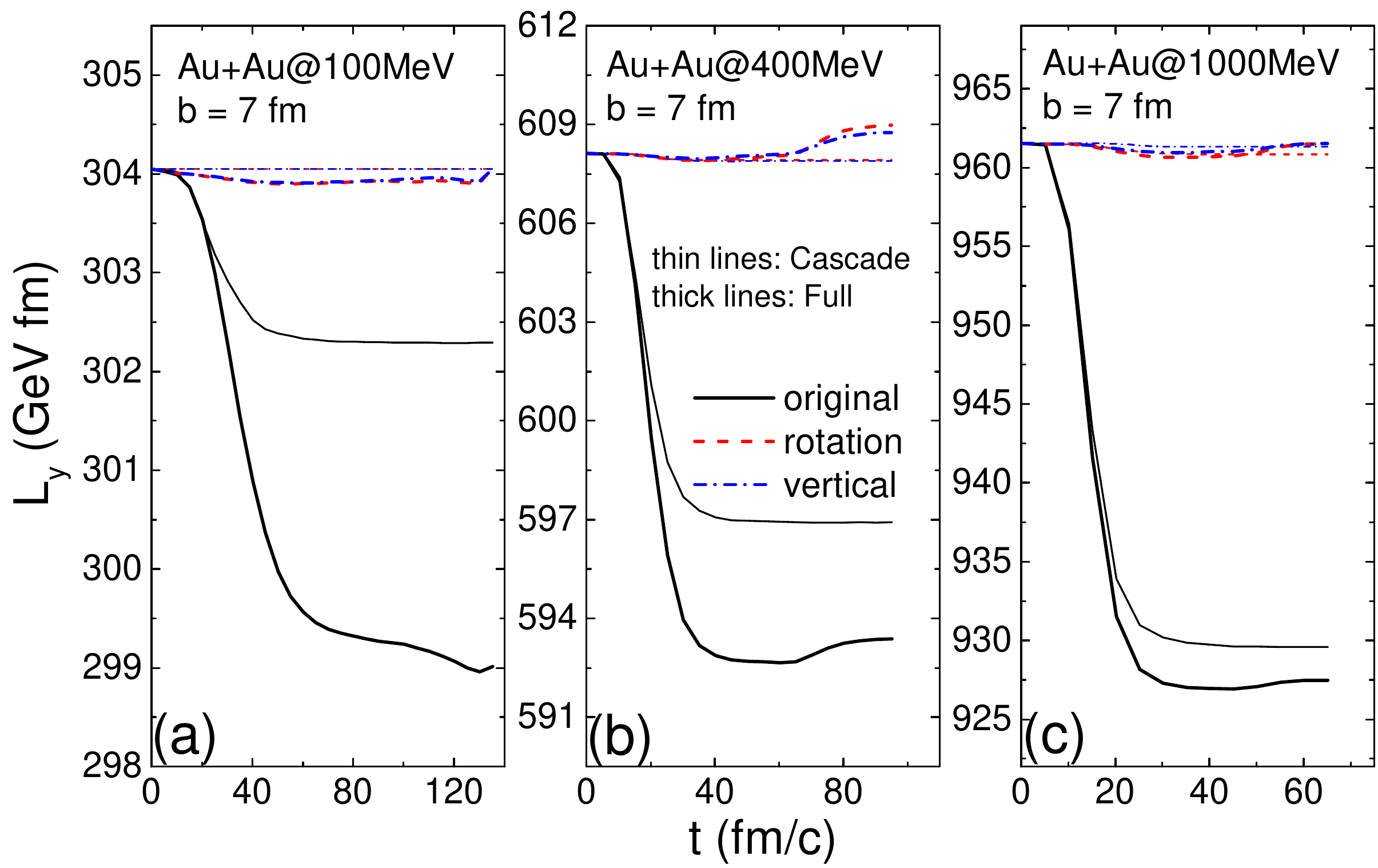}
\caption{\label{am} Time evolution of the total angular momentum perpendicular to the reaction plane from three collision prescriptions without (Cascade) and with (Full) nucleon mean-field potentials in non-central Au+Au collisions at $E_{lab}=100$ (a), 400 (b), and 1000 (c) AMeV. }
\end{figure}

We then move to simulations of non-central Au+Au collisions, and compare the total angular momentum from three collision prescriptions at different collision energies in Fig.~\ref{am}. In the Cascade mode with only elastic and inelastic collisions as well as Pauli blockings, the angular momentum $L_y$ perpendicular to the reaction plane is exactly conserved for the ``rotation'' and ``vertical'' prescriptions at 100 AMeV, and slightly violated at higher collision energies. The latter is due to the difficulty of conserving the angular momentum in the $N+\pi \rightarrow \Delta$ channel without incorporating spin degree of freedom, as mentioned before. The violation becomes even larger after incorporating the mean-field potential in the Full-mode calculation, but after all the violations are at the $0.1\%$ level. Incorporating the Coulomb potential between point charged particles in the present study always conserves the angular momentum. For the ``original'' collision prescription, although the angular momentum violation in one collision may cancel with that in another collision, the violation of the total angular momentum is much larger ($3\%$ at higher collision energies) compared to the ``rotation'' and ``vertical'' prescriptions. The violation of the total angular momentum in the ``original'' collision prescription is mostly due to nucleon-nucleon collisions at 400 and 1000 AMeV, but is seen to be much enhanced by the mean-field interaction at 100 AMeV.

\begin{figure}[ht]
\includegraphics[width=0.8\linewidth]{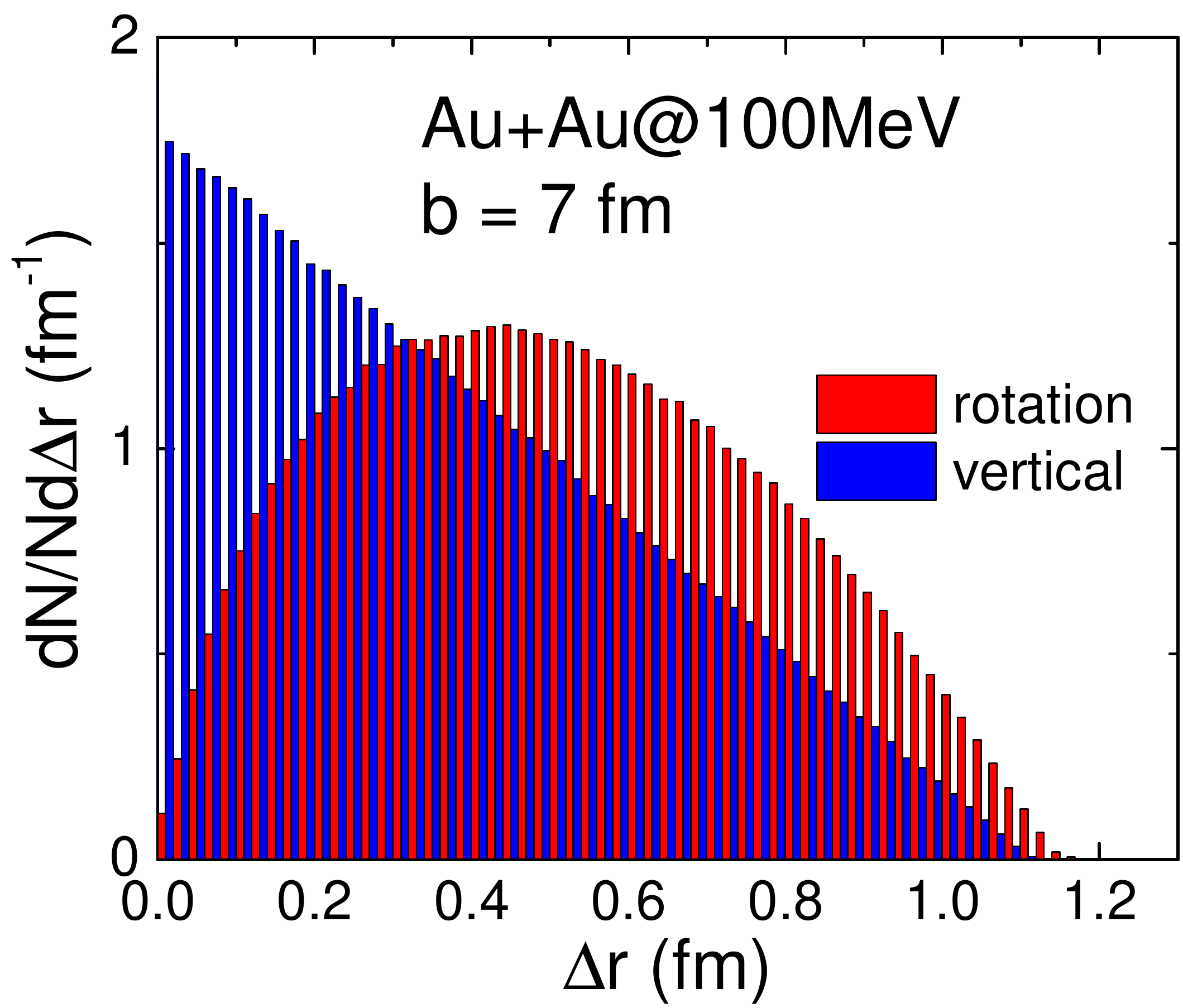}
\caption{Comparison of the histograms for the side-jump distance of particles in the computational frame from two collision prescriptions in non-central Au+Au collisions at $E_{lab}=100$ AMeV. The distributions are normalized to 1.}\label{fig2}
\end{figure}

The constraint of the angular momentum conservation generally requires the change of the coordinates of colliding particles after collision. Figure~\ref{fig2} compares the distribution of the side-jump distance $\Delta r$ in the computational frame from the ``rotation'' and ``vertical'' prescriptions in non-central Au+Au collisions at $E_{lab}=100$ AMeV, and the results are found to be insensitive to the collision energy. One sees that $\Delta r$ is smaller than 1.2 fm in all collisions, and is smaller in the ``vertical'' prescription than in the ``rotation'' prescription, even if the distance is now calculated in the computational frame.

\subsection{Nucleon dynamics}

We now compare the nucleon dynamics and nucleon observables in detail from three collision prescriptions in non-central Au+Au collisions. The comparisons are only in the Full-mode calculation with elastic and inelastic collisions as well as Pauli blockings, mean-field potential, and Coulomb potential.

\begin{figure}[ht]
\includegraphics[width=1\linewidth]{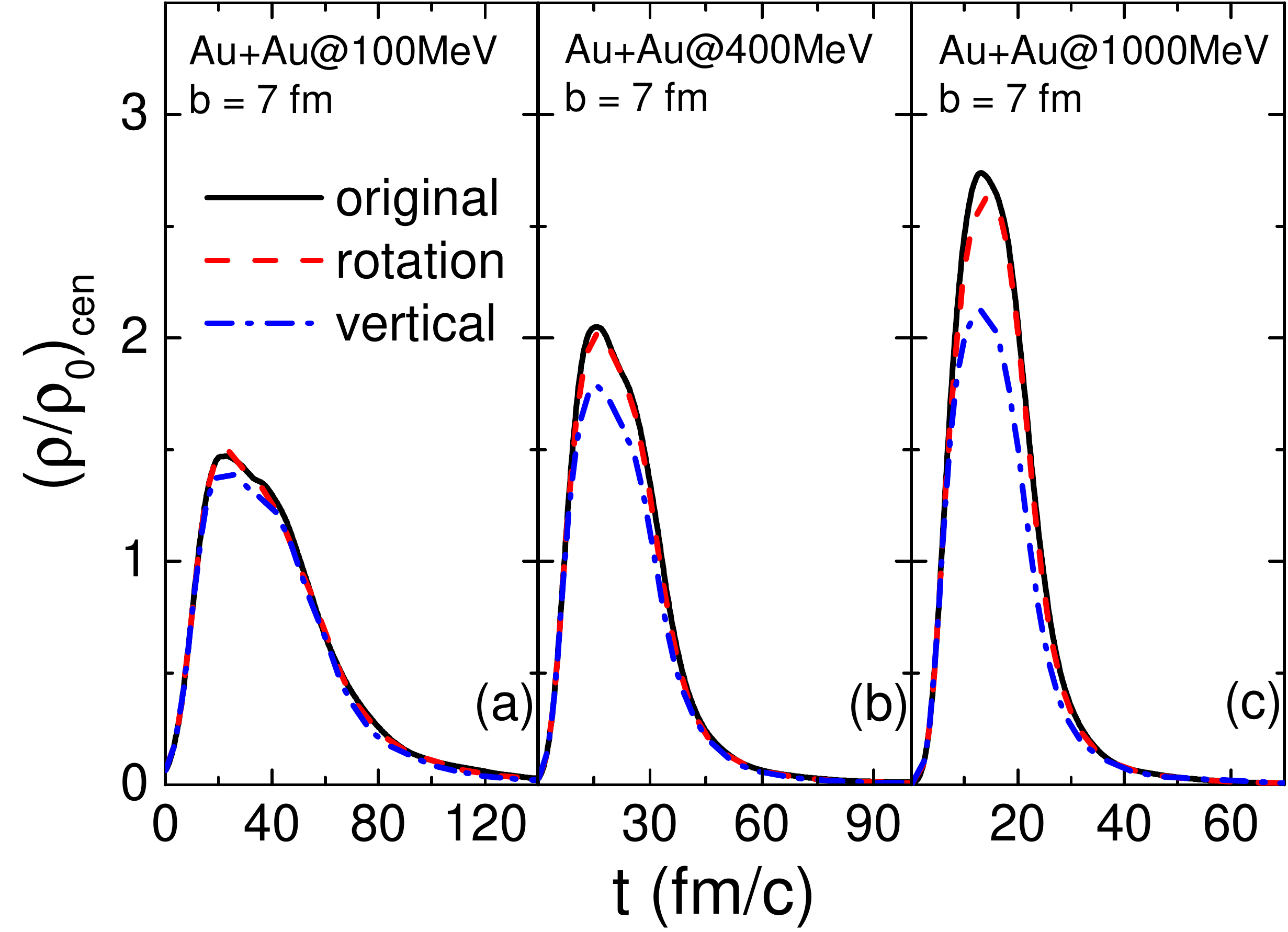}
\caption{\label{den} Time evolution of the central density from three collision prescriptions in non-central Au+Au collisions at $E_{lab}=100$ (a), 400 (b), and 1000 (c) AMeV.}
\end{figure}

Figure~\ref{den} compares the central density evolution from three collision prescriptions in non-central Au+Au collisions at different collision energies. As shown in Fig.~\ref{dNdt}, in-plane collisions, i.e., without changing the azimuthal angle after collisions, generally lead to a larger average distance between nucleons, thus a faster expansion of the system and a lower density. For the ``rotation'' prescription, the coordinates of the colliding particles are effectively pulled back a little after collisions, so this prescription leads to a slower expansion and a higher density compared to the ``vertical'' prescription, which happens to result in a central density evolution closer to that from the ``original'' prescription.

\begin{figure}[ht]
\includegraphics[width=1\linewidth]{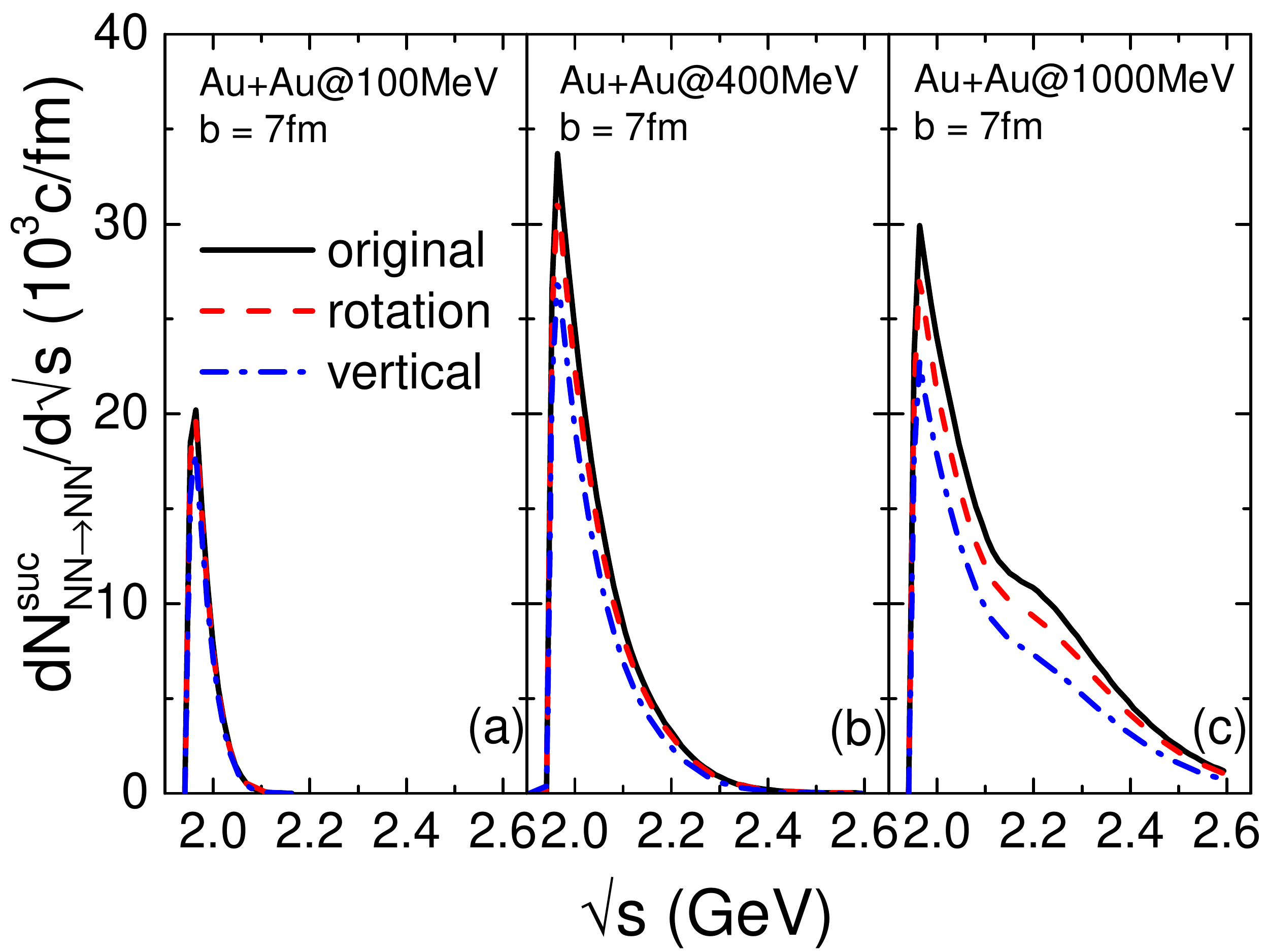}
\caption{\label{NNNN} C.M. energy dependence of the successful elastic nucleon-nucleon collision rate from three collision prescriptions in non-central Au+Au collisions at $E_{lab}=100$ (a), 400 (b), and 1000 (c) AMeV.}
\end{figure}

Figure~\ref{NNNN} compares the C.M. energy dependence of the successful elastic nucleon-nucleon collision rate from three collision prescriptions. With the same Pauli blocking implementation, the successful collision rate is dominated by the density evolution as shown in Fig.~\ref{den}. One sees that the ``vertical'' prescription leads to fewest collisions compared with other two prescriptions. The relative successful collision rates from the three prescriptions are qualitatively consistent with those in Fig.~\ref{dNdt} (a) where Pauli blockings are not implemented. We note that the correlation between the successful collision number and the dynamics becomes different in different collision prescriptions.

\begin{figure}[ht]
\includegraphics[width=1\linewidth]{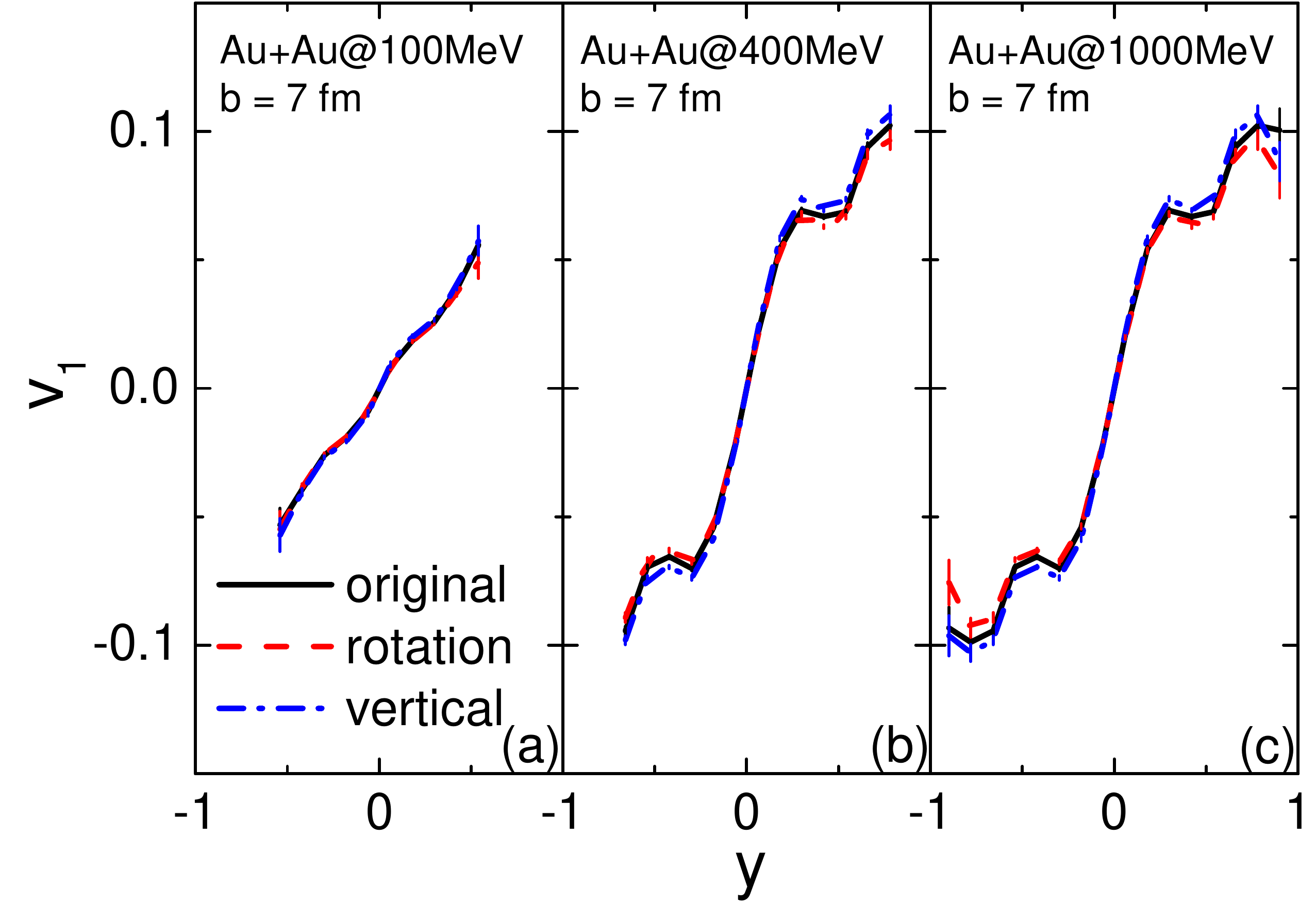}
\caption{\label{v1} Rapidity dependence of the directed flow of free nucleons from three collision prescriptions in non-central Au+Au collisions at $E_{lab}=100$ (a), 400 (b), and 1000 (c) AMeV.}
\end{figure}

\begin{figure}[ht]
\includegraphics[width=1\linewidth]{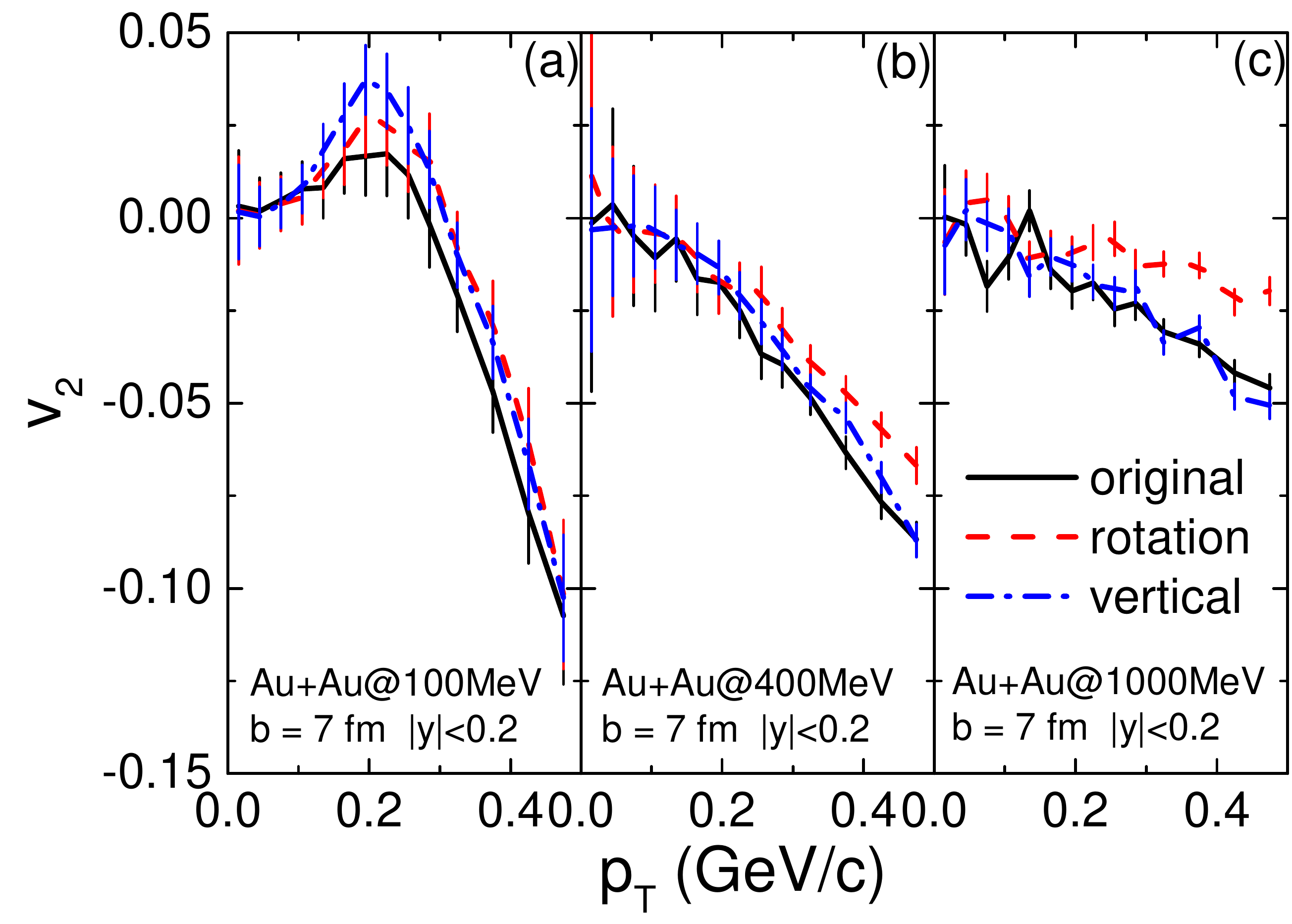}
\caption{\label{v2} Transverse momentum dependence of the elliptic flow of midrapidity ($|y|<0.2$) free nucleons from three collision prescriptions in non-central Au+Au collisions at $E_{lab}=100$ (a), 400 (b), and 1000 (c) AMeV.}
\end{figure}

Figures~\ref{v1} and \ref{v2} compares the rapidity dependence of the directed flow $v_1=\langle p_x/p_T \rangle $ and the transverse momentum dependence of the elliptic flow $v_2=\langle (p_x^2-p_y^2)/p_T^2 \rangle $ for free nucleons, which are determined by a cut-off density $\rho<0.15\rho_0$ at their freeze-out. $v_1$ from the three prescriptions are very similar, consistent with the smaller effect on the transverse flow found in Ref.~\cite{Gale:1990zz}, and the slightly stronger $v_1$ from the ``vertical'' prescription is likely due to the fast expansion of the system as mentioned above. The negative $v_2$ at large transverse momenta manifests the expansion of the participant matter blocked by the spectator matter and the squeeze-out of energetic nucleons perpendicular to the reaction plane. It seems that such squeeze-out effect is strongest from isotropic nucleon-nucleon collisions in the ``original'' prescription, since in non-central collisions nucleons are more likely to stay in the reaction plane with in-plane collisions. The ``rotation'' prescription has an effectively pulling-back movement, so nucleons are more likely to stay in the participant region. This further weakens the squeeze-out effect and makes the $v_2$ less negative at high transverse momenta, especially at higher collision energy.

For all results from Fig.~\ref{den} to Fig.~\ref{v2}, the difference among results from different collision prescriptions increases with increasing collision energy. This is expected since the Pauli blocking becomes less important and there are more successful nucleon-nucleon collisions at higher collision energies. This also shows the importance of incorporating properly the angular momentum conservation in the simulation of heavy-ion collisions at higher energies.

\subsection{Pion production}

We now compare with the pion productions from three collision prescriptions in non-central Au+Au collisions, by turning on the $N+N\leftrightarrow N+\Delta$ and $\Delta \leftrightarrow N+\pi$ channels for different isospin states. We neglect potentials for $\Delta$s and pions as well as the possible threshold effect in the present calculation, and only illustrate the effect of the angular momentum conservation in the collision prescription on the pion multiplicity and $\pi^-/\pi^+$ yield ratio.

\begin{figure}[ht]
\includegraphics[width=1\linewidth]{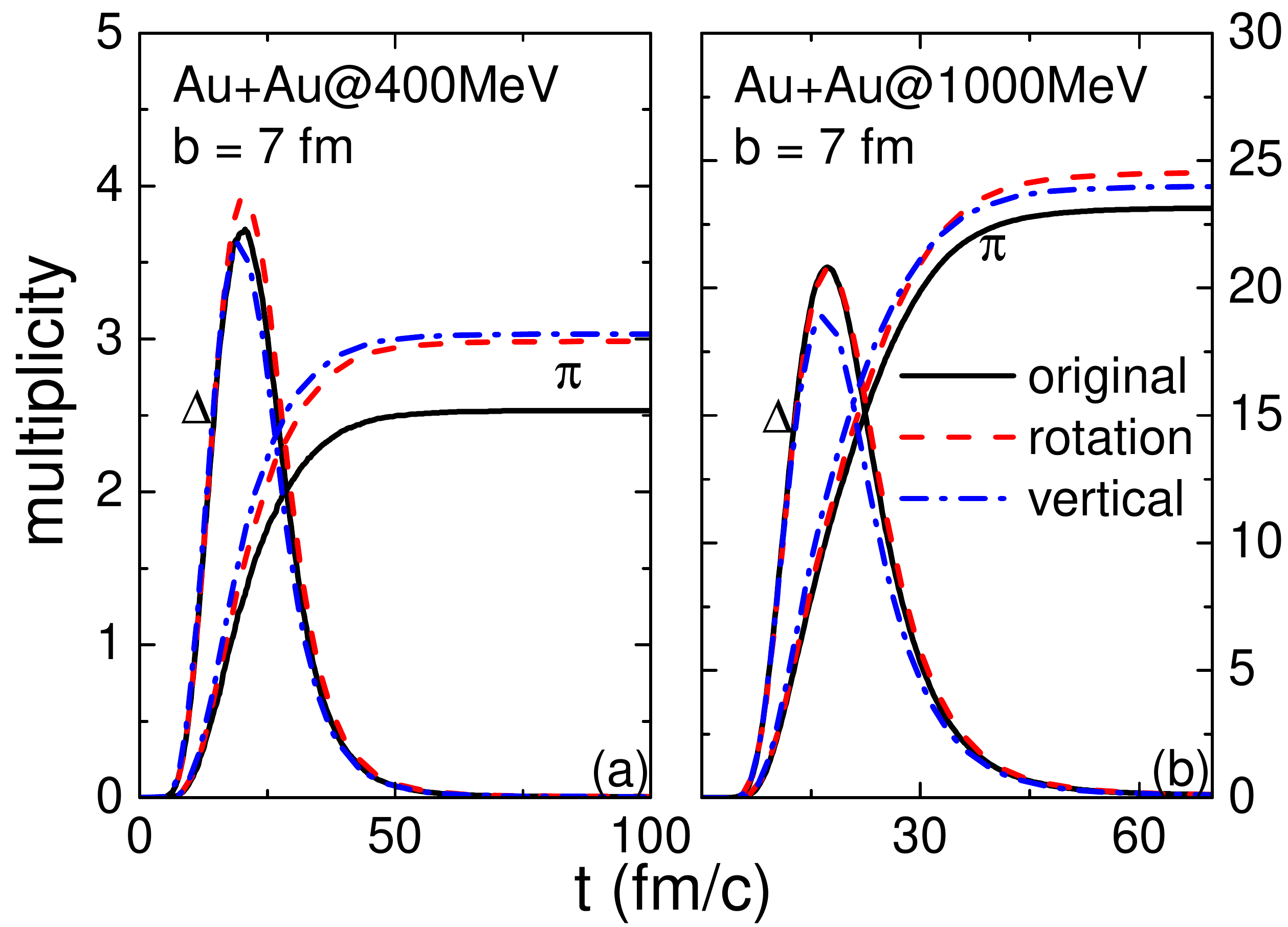}
\caption{\label{mul} Time evolution of the multiplicities of $\Delta$ resonances and pions from three collision prescriptions in non-central Au+Au collisions at $E_{lab}=400$ (a) and 1000 (b) AMeV.}
\end{figure}

Figure~\ref{mul} compares the time evolutions of the $\Delta$ and pion multiplicities from three collision prescriptions at different collision energies. The peak values of the $\Delta$ multiplicities in the three collision prescriptions are consistent with their maximum densities as shown in Fig.~\ref{den}. Compared with the ``original'' prescription, both channels of $N+N\rightarrow N+\Delta$ and $N+\Delta\rightarrow N+N$ are suppressed in the ``vertical'' prescription due to the larger $\Delta r_{min}$ as shown in Fig.~\ref{dNdt} (b). Since the $\Delta$ in the $N+\Delta\rightarrow N+N$ channel is farther from the nucleon in the ``vertical'' prescription compared to the ``original'' prescription, the $\Delta$ has a larger chance to decay. Therefore, the ``vertical'' prescription leads to a pion multiplicity that is about $20\%$ larger than that from the ``original'' prescription. Such effect is stronger at 400 AMeV but weaker at 1000 AMeV, due to the higher density and thus the weaker $\Delta r_{min}$ effect at higher collision energies. For the ``rotation'' prescription, there are more $\Delta$s produced compared with the ``vertical'' prescription, and the larger pion multiplicity than the ``original'' prescription can also be explained by the above $\Delta r_{min}$ effect.

\begin{figure}[ht]
\includegraphics[width=1\linewidth]{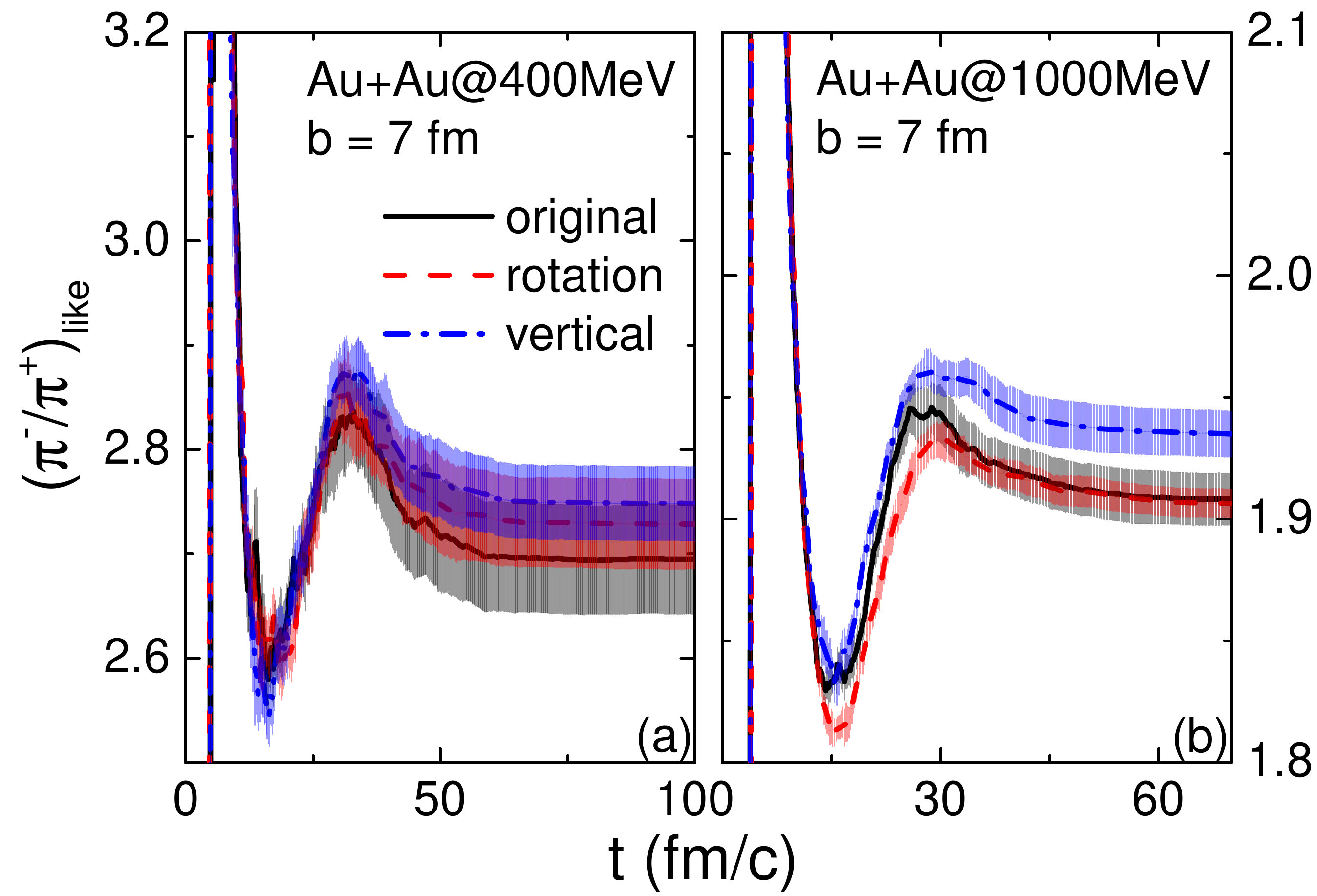}
\caption{\label{ratio} Time evolution of the $(\pi^-/\pi^+)_{like}$ ratio from three collision prescriptions in non-central Au+Au collisions at $E_{lab}=400$ (a) and 1000 (b) AMeV. }
\end{figure}

Figures~\ref{ratio} compares the time evolutions of the ratio $(\pi^-/\pi^+)_{like}=(\pi^-+\Delta^-+\frac{1}{3}\Delta^0)/(\pi^++\Delta^{++}+\frac{1}{3}\Delta^+)$ from three collision prescriptions at different collision energies. While the $(\pi^-/\pi^+)_{like}$ ratios are similar at $E_{lab}=400$ AMeV from different prescriptions within statistical error, the ``vertical'' prescription leads to a ratio which is about $1.5\%$ larger than those from the other two prescriptions at $E_{lab}=1000$ AMeV. The effect is smaller than that from the symmetry energy~\cite{Li:2002qx,Li:2002yda} but not negligible. The larger $(\pi^-/\pi^+)_{like}$ ratio from the ``vertical'' prescription is likely due to the stronger suppression on the $p+\Delta^- \rightarrow n+n$ channel as a result of less protons in the high-density region, compared with other prescriptions with smaller $\Delta r_{min}$ as shown in Fig.~\ref{dNdt} (b).

\section{Summary and outlook}
\label{sec:summary}

Based the framework of the IBUU transport model that is well calibrated by the previous efforts of the transport model evaluation project, we have revisited the dynamical effect of incorporating the rigorous angular momentum conservation with simple but reasonable homework setups. In order to conserve the angular momentum in each collision rigorously, the azimuthal angels of colliding particles should not be changed, while their coordinates should be adjusted, after each collision. We found that the option is not unique, and have compared results from two typical prescriptions, i.e., ``vertical'' and ``rotation'', with those from the ``original'' prescription. The ``vertical'' prescription requires the least change of the particle coordinates in the collision frame, and leads to a faster expansion of the system and a lower collision rate, and thus a lower central density and a slightly stronger directed flow in intermediate-energy heavy-ion collisions. This prescription also leads to a larger pion multiplicity and a larger $\pi^-/\pi^+$ yield ratio. The ``rotation'' prescription rotates the particle coordinates with respect to their center of mass, and leads to a similar but smaller effect as the ``vertical'' prescription, except that the ``rotation'' prescription results in a weaker elliptic flow.

The present study serves as an illustration of possible effects from the angular momentum conservation without introducing spin degree of freedom. In previous studies, we have investigated the spin dynamics~\cite{Xu:2015kxa} as well as the spin polarization~\cite{Xia:2019whr} induced by the spin-dependent potential in intermediate-energy heavy-ion collisions. It is of great interest to incorporate the constraint of total angular momentum conservation including the orbital contribution and the spin contribution, and to study the effect on the spin dynamics as well as the spin polarization. Such study is in progress.

\begin{acknowledgments}
We thank Che Ming Ko for helpful comments. This work is supported by the National Natural Science Foundation of China under Grant No. 11922514.
\end{acknowledgments}

\makeatletter
\providecommand \@ifxundefined [1]{%
 \@ifx{#1\undefined}
}%
\providecommand \@ifnum [1]{%
 \ifnum #1\expandafter \@firstoftwo
 \else \expandafter \@secondoftwo
 \fi
}%
\providecommand \@ifx [1]{%
 \ifx #1\expandafter \@firstoftwo
 \else \expandafter \@secondoftwo
 \fi
}%
\providecommand \natexlab [1]{#1}%
\providecommand \enquote  [1]{``#1''}%
\providecommand \bibnamefont  [1]{#1}%
\providecommand \bibfnamefont [1]{#1}%
\providecommand \citenamefont [1]{#1}%
\providecommand \href@noop [0]{\@secondoftwo}%
\providecommand \href [0]{\begingroup \@sanitize@url \@href}%
\providecommand \@href[1]{\@@startlink{#1}\@@href}%
\providecommand \@@href[1]{\endgroup#1\@@endlink}%
\providecommand \@sanitize@url [0]{\catcode `\\12\catcode `\$12\catcode
  `\&12\catcode `\#12\catcode `\^12\catcode `\_12\catcode `\%12\relax}%
\providecommand \@@startlink[1]{}%
\providecommand \@@endlink[0]{}%
\providecommand \url  [0]{\begingroup\@sanitize@url \@url }%
\providecommand \@url [1]{\endgroup\@href {#1}{\urlprefix }}%
\providecommand \urlprefix  [0]{URL }%
\providecommand \Eprint [0]{\href }%
\providecommand \doibase [0]{http://dx.doi.org/}%
\providecommand \selectlanguage [0]{\@gobble}%
\providecommand \bibinfo  [0]{\@secondoftwo}%
\providecommand \bibfield  [0]{\@secondoftwo}%
\providecommand \translation [1]{[#1]}%
\providecommand \BibitemOpen [0]{}%
\providecommand \bibitemStop [0]{}%
\providecommand \bibitemNoStop [0]{.\EOS\space}%
\providecommand \EOS [0]{\spacefactor3000\relax}%
\providecommand \BibitemShut  [1]{\csname bibitem#1\endcsname}%
\let\auto@bib@innerbib\@empty

%

\end{document}